\documentstyle[a4,12pt]{article}

\newcommand{\nit}{\noindent}

\newcommand{\np}{\newpage}
\newcommand{\dsp}{\displaystyle}
\newcommand{\vs}[1]{\vspace{#1 ex}}
\newcommand{\hs}[1]{\hspace{#1 em}}
\newcommand{\bfr}{\begin{flushright}}
\newcommand{\efr}{\end{flushright}}
\newcommand{\bc}{\begin{center}}
\newcommand{\ec}{\end{center}}
\newcommand{\ben}{\begin{enumerate}}
\newcommand{\een}{\end{enumerate}}

\newcommand{\be}{\begin{equation}}
\newcommand{\ee}{\end{equation}}
\newcommand{\ba}{\begin{array}}
\newcommand{\ea}{\end{array}}
\newcommand{\ct}{\cite}
\newcommand{\bit}{\bibitem}
\newcommand{\dd}[2]{\frac{\partial{#1}}{\partial{#2}}}
\newcommand{\ag}{\alpha}

\newcommand{\gam}{\gamma}
\newcommand{\del}{\delta}
\newcommand{\eps}{\epsilon}

\newcommand{\zg}{\zeta}
\newcommand{\thg}{\theta}
\newcommand{\vthg}{\vartheta} 

\newcommand{\lb}{\lambda}
\newcommand{\sg}{\sigma}
\newcommand{\rg}{\rho}
\newcommand{\fg}{\phi}
\newcommand{\ps}{\psi}
\newcommand{\vf}{\varphi}

\newcommand{\Gam}{\Gamma}
\newcommand{\Del}{\Delta}
\newcommand{\Fg}{\Phi}

\newcommand{\Og}{\Omega}

\newcommand{\beps}{\bar{\eps}} 
\newcommand{\bzg}{\bar{\zg}}

\newcommand{\bps}{\bar{\ps}}
\newcommand{\bxi}{\bar{\xi}}

\newcommand{\bFg}{\bar{\Fg}}

\newcommand{\bOg}{\bar{\Og}}

\newcommand{\bz}{\bar{z}} 
\newcommand{\bA}{\bar{A}}

\newcommand{\bF}{\bar{F}}

\newcommand{\ui}{\underline{i}}
\newcommand{\uj}{\underline{j}}

\newcommand{\uI}{\underline{I}}
\newcommand{\ta}{\tilde{a}}

\newcommand{\cL}{{\cal L}}

\newcommand{\lh}{\left(}
\newcommand{\rh}{\right)}

\newcommand{\slashed}{\hspace{-1.1ex}/}
\newcommand{\Slashed}{\hspace{-1.5ex}/\hspace{.2ex}}
\newcommand{\der}{\partial}
\newcommand{\Der}{D}
\newcommand{\sDer}{\Der\Slashed}
\newcommand{\sder}{\der\slashed}

\begin{document}

\pagestyle{empty}
\begin{flushright}
NIKHEF/03-012
\end{flushright} 
\vs{2} 

\begin{center} 
{\Large{\bf{K\"{a}hler manifolds and supersymmetry}}}\\
\vs{10}

{\large J.W.\ van Holten$^{*}$} \\
\vs{2}

{\large{NIKHEF, Amsterdam NL}} \\
\vs{3}

\today 
\vs{10}

{\small{ \bf{Abstract}}} 
\end{center}

\nit
{\footnotesize{Supersymmetric field theories of scalars and fermions in $4$-$D$ 
space-time can be cast in the formalism of K\"{a}hler geometry. In these lectures 
I review K\"{a}hler geometry and its application to the construction and analysis of 
supersymmetric models on K\"{a}hler coset manifolds. It is shown that anomalies 
can be eliminated by the introduction of  line-bundle representations of the coset 
symmetry groups. Such anomaly-free models can be gauged consistently and 
used to construct alternatives to the usual MSSM and supersymmetric GUTs. }} 

\vfill
\footnoterule 
\nit
{\footnotesize{
$^{*}$ \tt{e-mail: v.holten@nikhef.nl}} }

\np
~\hfill 

\np
\pagestyle{plain}
\pagenumbering{arabic}

\section{Supersymmetry}

Supersymmetry is a conjectured symmetry between the two fundamental classes
of particles observed in nature: bosons with integral spin, and fermions with odd
half-integral spin. The symmetry predicts bosons and fermions with the same mass
and the same quantum numbers (charges) in gauge intereactions. An important
motivation for the conjecture of supersymmetry is the fact that supersymmetry is
a direct  extension of the relativistic space-time symmetries described by the
Lorentz-Poincar\'{e} transformations; as such it has become a major component
of all viable models of quantum gravity, including supergravity and superstring theory.

The particle spectrum of the standard model, illustrated in table \ref{tab1}, does not
exhibit such a symmetry, certainly not in manifest form. Therefore it is necessary
to assume that supersymmetry is broken at energy scales of the standard model
and below, i.e.\ below 1 TeV.

At which energy above the Fermi scale supersymmetry is actually broken is model
dependent. If supersymmetry only plays a role in quantum gravity, it may well be
broken at the Planck scale ($10^{19}$ GeV).  Extrapolation of the running couplings
of the standard model indicates, that an approximately supersymmetric particle
spectrum at scales as low as the TeV scale would help to make the electro-weak
and color gauge couplings unify at an energy near $10^{15 \mbox{-} 16}$ GeV.
Supersymmetry breaking in the TeV range is the scenario underlying the minimal
supersymmetric standard model (MSSM) \ct{nilles}, in which all quarks and leptons
supposedly have scalar partners, and all gauge and  Higgs bosons (of which there
are at least two doublets) are accompanied by fermion partners, with appropriate mass
splittings largely adjusted by hand to fit observational constraints.

\begin{table}
\begin{center}
{\footnotesize
\begin{tabular}{|l|ccccc|} \hline
particle & color& isospin& isospin  &
 hypercharge & electric charge\\
 & multiplicity & multiplicity & $I_3$ & $Y$ & $Q = Y +I_3$ \\ \hline
 $\nu_L$ & 1 & 2 & +1/2 & $-1/2$ & ~~0 \\
 $e_L$   & 1 & 2 & $-1/2$ & $-1/2$ & ~$-1$ \\
 $\nu^c_L$ & 1 & 1 & 0 & 0 & ~~0 \\
 $e^c_L$  & 1 & 1 & 0 & $+1$ & ~+1 \\
 $u_L$  & 3 & 2 & $+1/2$ & $+1/6$ & ~2/3 \\
 $d_L$ & 3 & 2 & $-1/2$ & $+1/6$ & $-1/3$ \\
 $u^c_L$ & 3 & 1 & 0 & $-2/3$ & $-2/3$ \\
 $d^c_L$ & 3 & 1 & 0 & $+1/3$ & $+1/3$ \\ \hline
 $g$ & 8 & 1 & 0 & 0 & ~~0 \\
 $W^+$ & 1 & 3 & $+1$ & 0 & ~+1 \\
 $W^0$ & 1 & 3 & $0$ & 0 & ~~0 \\
 $W^-$ & 1 & 3 & $-1$ & 0 & ~$-1$ \\
 $B$ & 1 & 1 & 0 & 0 & ~~0 \\
 $H^+$ & 1 & 2 & $+1/2$ & $+1/2$ & ~+1\\
 $H^0$ & 1 & 2 & $-1/2$ & $+1/2$ & ~~0 \\ \hline
\end{tabular}
}
\caption{\footnotesize{Quantum numbers of quarks and leptons (upper part)
and bosons (lower part)  in the standard model.}}
\label{tab1}
\end{center}
\end{table}

More possibilities arise in models with large extra dimensions, such as that proposed
by Randall and Sundrum \ct{rand-sun}, which come naturally out of non-perturbat\-ive
string theory. In such models the Planck and unification scales are much closer to
the energy range of the standard model, and the constraints from gauge
unification are less stringent.

In the last 20 years much effort has been invested in the construction of supersymmetric
models with different particle spectra based on coset models, in which the coset $G/H$
is a K\"{a}hler manifold \ct{bagg-witt}-\ct{vh2}; for an early review, see \ct{bando-kugo-ya}.
The requirement of K\"{a}hler geometry, to be explained below, is natural in the context of 
$D = 4$ supersymmetry. Such models might arise as effective actions for low-energy
degrees of freedom, e.g.\ in strongly interacting supersymmetric gauge theories or 
composite models \ct{barnes}. From a string theory perspective they could be part of
an effective low-energy supergravity model; indeed, supergravity models often include 
non-linear coset models such as $SU(1,1)/U(1)$ in $D =4$, $N = 4$, and
$E_{7(+7)}/SU(8)$ in $D = 4$, $N = 8$ supergravity. 

A serious problem of supersymmetric models on K\"{a}hler cosets is, that they are
plagued by anomalies \ct{moor-nel}-\ct{buch-lerch}. To deal with this problem the
models have to be extended with additional superfields. One can for example
enlarge the symmetry group of the model, possibly with non-compact elements
\ct{ellw,buch-lerch}. In a direct bottom-up approach the anomalies can be canceled 
by additional supermultiplets carrying representations of the original coset $G/H$
\ct{sgn-vh1}.  However, this can only be done by including certain non-standard 
representations, first found in \ct{vh2}, in a novel way. In recent years models based
on this construction have been studied in great detail \ct{sgn-vh2, sgn-nyaw-vh,
sgn-nyaw-ricc-vh}, and we now have consistent supersymmetric models with
non-linear realizations of groups like $SU(5)$,  $SO(10)$, $E_6$ or $E_8$, and
new scenario's for superunification become possible.

The aim of these lectures is to present a pedestrian introduction to supersymmetric
coset models; they are organized as follows.  K\"{a}hler geometry and cosets are 
reviewed in a simple model providing insight in the abstract geometrical constructions.
The coupling of additional superfields to coset models of K\"{a}hler type is explained, 
as is their role in eliminating anomalies. I then turn to the more general formulation,
on which one can base more realistic models, including e.g.\ one or more generations
of quarks and leptons. I also discuss the effect of including gauge interactions. The
general methods are illustrated with the example of non-linear $SU(5)$.

\section{K\"{a}hler geometry: plane and sphere}

An $N$-dimensional complex manifold is a manifold which can be covered by a finite
set of local complex co-ordinate systems $(\bz^{\,\ui}, z^i)$, $(\ui, i) = 1,...,N$, such that
at the points at which the co-ordinate systems overlap the transition functions 
from one set of co-ordinates to the other are holomorphic:
\be 
\zg^i = f^i(z), \hs{2} \bzg^{\, \ui} = \bar{f}^{\,\ui}(\bz).
\label{2.0}
\ee
On such a manifold one can define a real line element of the form 
\be
ds^2 = g_{i\ui}(\bz, z)\, d\bz^{\ui}\, dz^i.
\label{2.0.1}
\ee
A complex manifold is a {\em K\"{a}hler} manifold if it satisfies the condition
that the holomorphic and anti-holomorphic curl of the metric vanishes:
\be
g_{i\ui, j} = g_{j\ui, i}, \hs{2} g_{i\ui,\uj} = g_{i\uj,\ui}.
\label{2.0.2}
\ee 
This condition can be written globally as the closure of a 2-form:
\be 
\der \Og = \bar{\der} \Og = 0, \hs{2} \Og = g_{i\ui}\,  d\bz^{\ui} \wedge dz^i.
\label{2.0.3}
\ee
Locally it implies that the metric can be derived from a real function 
$K(\bz,z)$ by
\be
g_{i\ui}(\bz,z) = \dd{^2 K}{z^i \der \bz^{\,\ui}}.
\label{2.0.4}
\ee
The function $K$ is called the {\em K\"{a}hler potential}. 

The simplest K\"{a}hler manifolds are the complex plane and the sphere. 
The line element in the plane is given everywhere by
\be
ds^2 = g_{z\bz}\, d\bz\, dz =  d\bz\, dz,
\label{2.1}
\ee
showing that there is a global real (hermitean) metric with a single component
$g_{z\bz} = 1$. This metric can be written as the mixed second derivative
of a real potential:
\be
g_{z\bz} = K_{,z\bz}, \hs{2} K(\bz,z) = \bz z. 
\label{2.3}
\ee
Therefore it automatically (though trivially) satisfies the curl condition (\ref{2.0.2}).

\begin{figure}
\let\picnaturalsize=N
\def\picsize{3.5in}
\def\picfilename{sphere.eps}
\ifx\nopictures Y\else{\ifx\epsfloaded Y\else\input epsf \fi
\let\epsfloaded=Y
\centerline{\ifx\picnaturalsize N\epsfxsize \picsize\fi \epsfbox{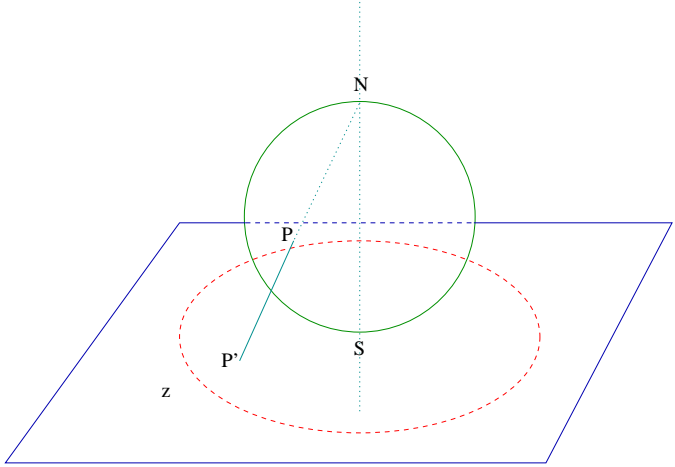}}}\fi
\caption{Map from the sphere to the tangent plane at the south pole.}
\label{fig1}
\end{figure}

The sphere can be covered locally by complex co-ordinates: the tangent space
at any point is the plane, parametrized by complex co-ordinates $(\bz, z)$. 
However, the map from the sphere to the tangent plane is not a global map: it
always excludes at least the point opposite that where the tangent plane is 
constructed. In figure \ref{fig1} the situation is sketched for the plane tangent
to the south pole of the sphere. 

The line element of the sphere with unit diameter can be translated from real polar 
co-ordinates, well-defined on the full sphere except for the poles, to complex
co-ordinates in the tangent plane by 
\be
\ba{l}
ds^2\, =\, \dsp{ R^2 \lh d\vthg^2 + \sin^2 \vthg\, d\vf^2 \rh } \\
  \\
  \dsp{\hs{1.2} \stackrel{R = 1/2}{=}\, \frac{d\bz dz}{(1 + \bz z)^2} = g_{z\bz}\, d\bz\, dz. }
\ea
\label{2.4}
\ee

\vs{1}

\nit
The hermitean metric is obtained from a real K\"{a}hler potential
\be
g_{z\bz} = K_{,z\bz}, \hs{2}
K(\bz,z) = \ln (1 + \bz z)
\label{2.5}
\ee
The complex co-ordinates cover the whole sphere minus the north pole $N$.
Observe, that this particular projection maps the equator to the unit circle, the
southern hemisphere to its interior, and the northern hemisphere to its exterior.

\begin{figure}
\let\picnaturalsize=N
\def\picsize{3.5in}
\def\picfilename{Nsphere.eps}
\ifx\nopictures Y\else{\ifx\epsfloaded Y\else\input epsf \fi
\let\epsfloaded=Y
\centerline{\ifx\picnaturalsize N\epsfxsize \picsize\fi \epsfbox{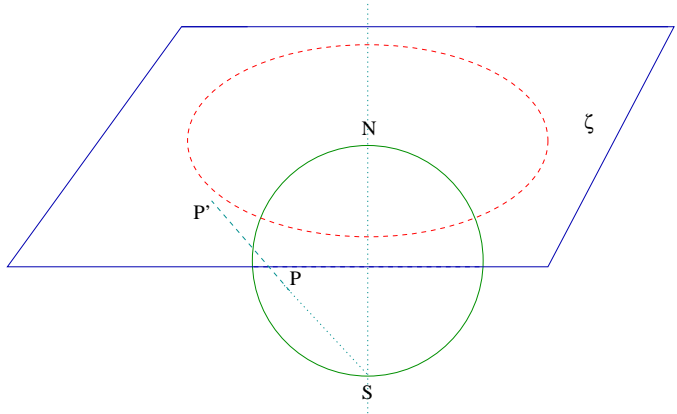}}}\fi
\caption{Map from the sphere to the tangent plane at the north pole.}
\label{fig2}
\end{figure}

A second map including the north pole is obtained by inversion of the co-ordinates,
i.e.\ the holomorphic co-ordinate transformation
\be
\zg = \frac{1}{z}, \hs{2} \bzg = \frac{1}{\bz}.
\label{2.6}
\ee
The $\zg$-plane is the plane tangent to the north pole; it defines a complex
co-ordinate system covering the sphere minus the south pole, as in figure \ref{fig2}.
This is easily observed, as again the equator is projected onto the unit circle, but
now the southern hemisphere is mapped to the exterior, and the northern hemisphere
is mapped to the interior;  in particular the north pole corresponds to the
origin $\zg = 0$.
Furthermore, the inversion does not change the expression for the line element.
This is easy to understand on the basis of symmetry: no matter where the
tangent plane is constructed, the spherical symmetry implies that the line
element will always be of the form
\be
ds^2 = \frac{d\bar{\zg} d\zg}{(1 + \bzg \zg)^2}.
\label{2.7}
\ee
Note, that inversion changes the K\"{a}hler potential to
\be
K[\bz(\bzg), z(\zg)] = \ln \lh1 + \frac{1}{\bzg \zg} \rh =
 \ln \lh 1 + \bzg \zg \rh + F(\zg) + \bF(\bzg),
\label{2.8}
\ee

\vs{1}

\nit
where $F(\zg) = - \ln \zg$ modulo an arbitrary imaginary constant. It follows immediately,
that
\be
g_{,\zg \bzg} = K_{,\zg\bzg} = \frac{1}{(1 + \bzg  \zg)^2},
\label{2.9}
\ee
as expected.

\section{Symmetries of the sphere}

The sphere is by definition invariant under rotations around  three independent
axes. We consider an arbitrary infinitesimal rotation (a rotation close to the identity)
over angles $(\vf, \vthg, \eta)$. The corresponding change in projection of a rotated
point to the tangent plane corresponds to a change in the complex co-ordinates
\be
z^{\prime} = z + \del z, \hs{2}
\del z = i \vf z + \frac{\vthg}{2} \lh 1 + z^2 \rh + \frac{i\eta}{2} \lh 1 - z^2 \rh.
\label{3.1}
\ee
This is a special {\em holomorphic} co-ordinate transformation which leaves the
line element of the sphere invariant:
\be
\frac{d\bz^{\,\prime} dz^{\prime}}{(1+ \bz^{\,\prime} z^{\prime})^2}
 = \frac{d\bz dz}{(1 + \bz z)^2}.
\label{3.2}
\ee
However, again the K\"{a}hler potential is invariant only modulo the real 
part of a holomorphic function:
\be
K^{\prime}(\bz^{\,\prime}, z^{\prime}) = \ln (1 + \bz^{\,\prime} z^{\prime})
 = K(\bz, z) + f(z) + \bar{f}(\bz),
\label{3.3}
\ee
where the holomorphic function is given by
\be 
f(z) = \frac{1}{2} \lh i \vf +(\vthg - i\eta) z \rh.
\label{3.4}
\ee
Actually, the purely imaginary $z$-independent part of $f(z)$ is not determined by the 
transformation of the K\"{a}hler potential. We have chosen this particular expression
for later convenience. Note, that in general invariance of the K\"{a}hler potential modulo 
the real part of a holomorphic function is sufficient for invariance of the line element:
\be 
g^{\prime}_{z^{\prime} \bz^{\prime}}(\bz^{\,\prime}, z^{\prime})
 = \dd{^2K^{\prime}}{z^{\prime} \der \bz^{\,\prime}}\,
 = \frac{1}{\lh 1 + \bz^{\,\prime} z^{\prime} \rh^2} = g_{z\bz}(\bz^{\prime}, z^{\prime}).
\label{3.5}
\ee
A vector field $\xi(z) = \del z$ which, like (\ref{3.1}), defines an infinitesimal co-ordinate 
transformation leaving the line element invariant, is called a Killing vector.
The idea carries over directly to higher-dimensional manifolds: on an $N$-dimensional 
complex manifold a (holomorphic) Killing vector is a (holomorphic) transformation
\be
z^{\prime\, i} = z^i + \xi^i(z), \hs{2}
\bz^{\prime\,\ui} = \bz^i + \bar{\xi}^{\,\ui}(\bz),
\label{3.6}
\ee
such that the line element is invariant: 
\be
ds^2 = g_{i\ui}(\bz,z)\, d\bz^{\ui}\, dz^i = g_{i\ui}(\bz^{\prime}, z^{\prime})\, 
 d\bz^{\prime\,\ui}\, dz^{\prime\,i}.
\label{3.7}
\ee
This happens, if the transformation $(\bar{\xi}^{\,\ui}(\bz), \xi^i(z))$ satisfies the 
condition
\be
\dd{\xi_{\ui}}{z^i} + \dd{\bar{\xi}_{i}}{\bz^{\ui}} = 0,
\label{3.8}
\ee
where $\xi_{\ui} = g_{i\ui}\, \xi^i$ and $\bar{\xi}_i = g_{i\ui}\, \bar{\xi}^{\ui}$. We 
will encounter more examples of such Killing vector fields later on.

\section{Cosets}

There is another way of looking at the sphere, in terms of groups and cosets.
In this section we describe how the sphere can be identified with the coset
$SU(2)/U(1)$.

Consider an element $a$ of $SU(2)$. By definition it is a unitary $2 \times 2$
matrix with unit determinant: $a a^{\dagger} = 1$, $\det a = 1$. All elements in
the neighborhood of the identity can be parametrized in terms of a real
parameter $\ag$ and a complex parameter $z$ by
\be
a(\bz,z;\ag) = \frac{e^{i\ag \tau_3}}{\sqrt{1 + \bz z}}\, \lh \ba{cc} 1 & z \\
                                                                                     -\bz & 1 \ea \rh.
\label{4.1}
\ee
Note, that the complete $\ag$-dependence is in the $U(1)$ factor
$\exp (i \ag \tau_3)$. The class of all elements corresponding to the same
value of $(\bz,z)$ is therefore a set of elements of $SU(2)$ differing only 
by a $U(1)$ factor; the set of all such equivalence classes is the coset
$SU(2)/U(1)$. A representative of each equivalence class can be chosen
by fixing the $U(1)$ gauge to $\ag = 0$:
\be
\ta (\bz,z) = a(\bz,z;0) = \frac{1}{\sqrt{1 + \bz z}} \lh \ba{cc} 1 & z \\
                                                                                           -\bz & 1 \ea \rh.
\label{4.2}
\ee
Thus we can associate one element of the coset (an element in the neighborhood
of the identity) represented by $\ta$ with each point in the complex plane. However, 
there is one element of the coset which is not in the neighborhood of the identity,
as it is not associated with any point in the (finite) plane: the element
\be
\ta_{\infty} = \lh \ba{cc} 0 & 1\\
                                    -1 & 0 \ea \rh.
\label{4.3}
\ee
In the present approach  this element can be reached by an inversion
of the co-ordinates: $\zg = 1/z$, and a change of gauge by multiplying with the $U(1)$
element $\exp (i \thg \tau_3)$, where $\thg = \arg z = - \arg \zg$:
\be
a(\bzg^{-1}, \zg^{-1}; \thg) = \frac{1}{\sqrt{1 + \bzg \zg}} 
 \lh \ba{cc} e^{-i\thg} \sqrt{\bzg \zg} & 1 \\
                 -1 & e^{i\thg} \sqrt{\bzg \zg} \ea \rh = \frac{1}{\sqrt{1 + \bzg \zg}} 
 \lh \ba{cc} \zg & 1 \\
                 -1 & \bzg \ea \rh,
\label{4.4}
\ee
It is then obvious that
\be
\ta_{\infty} = \lim_{\zg \rightarrow 0} a(\bzg^{-1}, \zg^{-1}; \thg).
\label{4.5}
\ee
Thus the elements of the coset are actually in one-to-one correspondence 
with the points on the sphere, rather than with the points in the complex
plane. \vs{1}

\nit
Not only is there a one-on-one map between coset $SU(2)/U(1)$ and the
sphere, this map also respects the symmetries of the sphere, as we now
explain. Consider any element $\ta(\bz,z)$ of the coset in the gauge $\ag = 0$.
By construction it is also an element of $SU(2)$. We can multiply $\ta$
from the right with an other element $g$ of $SU(2)$, to get the new element
$a_g = \ta g$. In general this will not be an element satisfying the gauge
condition $\ag = 0$. It is possible to get back to this gauge by multiplying the
element $\ag_g$ from the left by a $g$- and point-dependent $U(1)$ transformation 
$h_g(\ta)$, such that
\be
\ta_g = h_g \ta g =\frac{1}{\sqrt{1 + \bz_g z_g}} \lh \ba{cc} 1 & z_g \\
                                                                                           -\bz_g & 1 \ea \rh.
\label{4.6}
\ee
In this way any $SU(2)$ transformation $g$ maps a coset element $\ta$ labeled by
parameters $(\bz, z)$ to another element $\ta_g$ labeled by $(\bz_g, z_g)$. 
Explicitly, for an infinitesimal transformation
\be 
g \simeq \lh \ba{cc} 1 - i \vf/2 & (\vthg + i \eta)/2 \\
                      - (\vthg - i \eta)/2 & 1 + i \vf/2 \ea \rh,
\label{4.7}
\ee
there is a compensating $U(1)$ transformation
\be 
h_g(\bz, z) \simeq \lh \ba{cc} 1 + i \ag(\bz, z) & 0 \\
                                              0 & 1 - i \ag(\bz, z) \ea \rh, 
\label{4.8}
\ee
with
\be 
\ag(\bz, z) = \frac{1}{2i} \lh f(z) - \bar{f}(\bz) \rh,
\label{4.9}
\ee
such that $\ta_g$ is given by the r.h.s.\ of (\ref{4.6}) with
\be
z_g \simeq  z + \del z, \hs{2} 
\del z = i \vf z + \frac{\vthg}{2} \lh 1 + z^2 \rh + \frac{i\eta}{2} \lh 1 - z^2 \rh.
\label{4.10}
\ee
as in eq.(\ref{3.1}); $f(z)$ in eq.(\ref{4.9}) is given by the expression (\ref{3.4}),  
which justifies {\em a posteriori} our choice of the constant imaginary term in
$f(z)$. In conclusion, we see that the symmetries of the sphere (rotations)
are realized as non-linear $SU(2)$ transformations on the coset elements
$\ta(\bz, z)$. Therefore, the invariant line element on the sphere (\ref{2.4}) also 
defines an $SU(2)$-invariant line element on the coset $SU(2)/U(1)$.

\section{Field theory on cosets: the non-linear $\sg$-model}

We have seen how a non-linear realization of $SU(2)$ can be constructed in
terms of coset elements. We have also constructed an invariant line element 
on the coset. It is then straightforward to write down a theory of a massless complex
scalar field $z(x)$ in a $n$-dimensional space-time, which is invariant under the 
non-linear $SU(2)$ transformations (\ref{4.10}).  It is based on the invariant action
\be 
S[\bz, z] = - \int d^n x\, \frac{\der \bz \cdot \der z}{(1 + \bz z)^2}.
\label{5.1}
\ee
As our discussions above show, this action can only be used for
fluctuations of the field not too far from the identity $(\bz = z = 0)$. For large
fluctuations which include the pole of the sphere at $z = \infty$, a second 
chart from the sphere to the complex plane is needed. Another limitation is,
that the model is not renormalizable beyond $n = 2$; therefore in $4$-$d$ 
space-time such a field theory is at best an effective theory for light scalar
degrees of freedom in a theory with broken $SU(2)$. 

It is known from Goldstone's theorem, that in fact such massless scalars always
arise in a theory with a spontaneously broken rigid symmetry, either as elementary
or composite states. For the case of elementary fields it is easy to show. Suppose
that we have a triplet of (real) scalar fields $\fg_i$, $i = (1,2,3)$, with an action
\be
S[\vec{\fg}] = - \int d^n x\, \lh \frac{1}{2}\, (\der \vec{\fg})^2 + V[\vec{\fg}^{\,2}] \rh.
\label{5.2}
\ee
As the potential depends only on the modulus $\vec{\fg}^{\,2}$, the action is invariant 
under $SO(3) \simeq SU(2)$ transformations rotating the 3-vector $\vec{\fg}$ whilst
keeping its length fixed. Suppose the lowest energy state of the theory is such 
that the field has a non-zero expecation value: $\vec{\fg}^{\,2} = 1/f^2$. Then the
low-energy behaviour of the theory is dominated by the fluctuations of $\vec{\fg}$
respecting the fixed length; these can be parametrized by writing
\be 
\ba{l}
\dsp{ \fg_1 = \rg\, \sin \vthg \cos \vf, \hs{2} \fg_2 = \rg\, \sin \vthg \sin \vf, }\\
 \\
\dsp{ \fg_3 = \rg\, \cos \vthg. }
\ea
\label{5.3}
\ee
Taking $\rg = 1/f$ fixed, and inserting this parametrization back into the action 
(\ref{5.2}), we obtain
\be 
S[\vec{\fg}] \rightarrow \frac{1}{f^2}\, \int d^n x\, \left[ (\der \vthg)^2
 + \sin^2 \vthg\, (\der \vf)^2 + V[1/f^2] \right]. 
\label{5.4}
\ee
A comparison with eq.(\ref{2.4}) shows that up to an additive constant $V[1/f^2]$,
and with the normalization $f = 2$, this reduces to the action (\ref{5.1}). On the other
hand, keeping the radial degree of freedom $\rg$ in the theory, one
easily deduces that it has a mass $m_{\rg}^2 = \der^2 V/ \der \rg^2$
(evaluated at $\rg=1/f$). This establishes that the range of energies, in which the
effective action (\ref{5.1}) is valid, is $-Q^2 \leq m_{\rg}^2$. If the symmetry
discussed is broken at very high energies, such as in GUT models, this
regime can of course be very large ($m_{\rg} \sim 10^{15}$ GeV).

\section{Dressing the sphere}

In the following we extend this construction to supersymmetric field theories.
To this end we have to include additional fields also transforming under some
non-linear (spontaneously broken) version of $SU(2)$, or some larger Lie group 
in more general cases. The key to finding representations of the non-linear
transformations is by their identification as special holomorphic co-ordinate 
transformations. This implies, that if we have representations of the group of
general co-ordinate transformations, we directly obtain representations of 
$SU(2)$ by restricting the co-ordinate transformations to those generated by
the Killing vectors (\ref{3.1}).

The simplest representations we can construct are those based on standard 
representations of the group of general co-ordinate transformations:
scalars, vectors and tensors. A scalar $S(\bz,z)$ takes the same value at
the same point, independent of the co-ordinate system; hence under an
infinitesimal holomorphic co-ordinate transformation
\be
z^{\prime} = z + \xi(z), \hs{2} \bz^{\prime} = \bz + \bar{\xi}(\bz), 
\label{6.1}
\ee
the scalar transforms to first order in $(\bar{\xi}, \xi)$ as
\be 
S^{\prime}(\bz^{\prime}, z^{\prime}) = S(\bz,z) \hs{1} \Rightarrow \hs{1}
S^{\prime}(\bz, z) - S(\bz, z) =  - \xi(z)\, \der S(\bz, z) - \bar{\xi}(\bz)\, \bar{\der} S(\bz, z).
\label{6.2}
\ee
Applying the same rule to the components of a holomorphic one-form
$A(\bz, z)$ $= A_z(\bz, z)\, dz$ one finds:
\be
A^{\prime}_z (\bz, z) - A_z(\bz, z) = - \der \xi(z) A_z(\bz, z) - \xi(z)\, \der A_z(\bz, z)
 - \bar{\xi}(\bz)\, \bar{\der} A_z(\bz, z).
\label{6.3}
\ee
For the components of a vector $V(\bz, z) = V^z(\bz, z)\, \der$ we find the 
contragredient transformation:
\be 
V^{\prime\, z}(\bz, z) - V^z(\bz, z) = \der \xi(z) V^z(\bz, z)
 - \xi(z)\, \der V^z(\bz, z) - \bar{\xi}(\bz)\, \bar{\der} V^z (\bz, z).
\label{6.4}
\ee
This guarantees that the contraction of a vector and a one-form of the
same type (e.g., $V^z A_z$) transforms as a scalar. In addition to holomorphic
one-forms and vectors, there are also anti-holomorphic one-forms and vectors
($\bar{A} = \bar{A}_{\bz}\, d\bz$, $\bar{V} = \bar{V}^{\, \bz}\, \bar{\der}$), 
and tensors transforming as direct products of forms and vectors. e.g.:
\be 
\ba{lll}
T^{\prime}_{z\bz}(\bz, z) - T_{z\bz}(\bz, z) & = & \dsp{  
 - \der \xi(z) T_{z\bz}(\bz, z) - \bar{\der} \bar{\xi} (\bz) T_{z\bz}(\bz, z) }\\
 & & \\
 & & \dsp{ - \xi(z) \der T_{z\bz}(\bz, z) - \bar{\xi}(\bz) \bar{\der} T_{z\bz}(\bz, z) }\\
 & & \\
 & = & \dsp{ - \der \lh \xi(z) T_{z\bz}(\bz, z) \rh - \bar{\der} \lh \bar{\xi}(\bz)
 T_{z\bz}(\bz, z) \rh. } 
\ea
\label{6.5}
\ee
In particular, the metric transforms as such a mixed tensor, from which one
can immediately deduce the Killing condition (\ref{3.8}) for invariance of the line
element.

Each of the transformations (\ref{6.2})-(\ref{6.5}) can be turned into an
$SU(2)$ transformation corresponding to the element $g(\vf, \vthg, \eta)$ by
restricting $(\xi(z), \bar{\xi}(\bz))$ to the Killing vectors (\ref{3.1}):
\[
\del_g z = \xi_g(z) = i \vf z + \frac{\vthg}{2} \lh 1 + z^2 \rh + \frac{i\eta}{2} \lh 1 - z^2 \rh.
\]
{\em Line bundles} \\ 
In addition to the vector and tensor representations constructed in this way,we can
define representations based on holomorphic line bundles. The construction starts
from the transformation rule of the K\"{a}hler potential: $\del_{\xi} K = F_{\xi} + \bF_{\xi}$,
where $\xi$ is a Killing vector and $F_{\xi}$ is a corresponding holomorphic function. A
holomorphic line bundle $\Og(\bz, z)$ of weight $\lb$, defined on the sphere, then transforms
under a holomorphic co-ordinate transformation (\ref{6.1}) as
\be
\ba{l}
\Og^{\prime}(\bz^{\prime}, z^{\prime})\, =\, \dsp{ e^{\lb F_{\xi}(z)}\, \Og(\bz, z) }\\
 \\
 ~~\Rightarrow\; \dsp{  \Og^{\prime}(\bz, z) - \Og(\bz, z) \approx
 - \xi(z)\, \der \Og - \bxi(\bz)\, \bar{\der} \Og + \lb F_{\xi}(z) \Og(\bz, z). }
\ea
\label{6.6.0.0}
\ee
In particular, under an $SU(2)$ transformation defined by the
above Killing vector:
\be
\Og_g(\bz_g, z_g) = e^{\lb f_g(z)} \Og(\bz, z)
 \approx \Og(\bz, z) + \lb f_g(z) \Og(\bz, z). 
\label{6.6.0}
\ee
This defines a representation of $SU(2)$, as the functions $f_g(z)$ have
the property 
\be
 f_{g_2 g_1}(z_{g_2 g_1}) = f_{g_2}(z_{g_1}) + f_{g_1} (z).  
\label{6.6.1}
\ee
Similarly, one can define anti-holomorphic line bundles $\bOg(\bz, z)$
with the transformation law
\be
\bOg_g (\bz_g, z_g) = e^{\lb \bar{f}_g(\bz)}\, \bOg(\bz, z)
 \approx \bOg(\bz, z) + \lb \bar{f}(\bz) \Og(\bz, z).
\label{6.6.2}
\ee
The archetype of a multiplicative line bundle on a K\"{a}hler manifold is the
exponent of the K\"{a}hler potential:
\be 
E_{\lb}(\bz, z) = e^{\lb K(\bz,z)}.
\label{6.6}
\ee
Under an $SU(2)$ transformation $g(\vf, \vthg, \eta)$ of the type above,
cf.\ eqs.\ (\ref{4.7})-(\ref{4.10}), $E_{\lb}$ transforms as
\be 
\ba{lll}
(E_{\lb})_g(\bz_g, z_g) & = & \dsp{
 e^{\lb f_g(z) + \lb \bar{f}_g(\bz)}\, E_{\lb}(\bz ,z) }\\
 & & \\
 & \approx & \dsp{ E_{\lb}(\bz, z) + \lb f_g(z) E_{\lb}(\bz, z)
    + \lb \bar{f}_g(\bz) E_{\lb}(\bz ,z). }
\ea
\label{6.7}
\ee
The condition (\ref{6.6.1}) is sufficient to guarantee the group property of the 
transformations for a line bundle $\Og(\bz, z)$ in any point of the manifold.
However, it does not guarantee the existence of the line bundle 
globally; the global existence of line bundles requires $\lb \in {\bf Z}$. Indeed,
under the transformation $z^{\prime} = 1/z$, necessary to cover the full coset (sphere) 
\be
\Og^{\prime} (\bz^{\prime}, z^{\prime}) = \frac{1}{z^{\lb}}\, \Og(\bz, z),
\label{6.8}
\ee
This is single-valued on the unit circle (the equator of the sphere, where the
co-ordinate patches overlap) if and only if $\lb$ is an integer. A similar
quantization condition (the {\em cocycle condition}) holds on all compact
K\"{a}hler cosets. 

Finally, combining the various realizations of the non-linear symmetry transformations
constructed here, the most general representation of spontaneously broken $SU(2)$
is a combination of vector/tensor- and line-bundles, e.g.\ a one-form valued mixed
holomorphic and anti-holomorphic line bundle with component $A_z$ transforms as
\be
\ba{l}
A_z^{\prime}(\bz, z) - A_z(\bz, z) = \dsp{ -\, \der \xi_g(z)\, A_z(\bz, z)
 - \xi_g(z) \der A_z(\bz, z) - \bxi_g(\bz)\, \bar{\der}  A_z(\bz, z) }\\
  \\
 \dsp{ \hs{9.4} +\, \lh \lb f_g(z) + \mu \bar{f}_g(\bz) \rh A_z(\bz, z), }
\ea
\label{6.9}
\ee
where $\lb$ and $\mu$ are integers for global consistency.

\section{Supersymmetry on the sphere}

With the prescriptions of the previous section at hand we can now construct
a theory with supersymmetry in the target space defined by the sphere --- 
or some other K\"{a}hler manifold. This construction goes as follows. In $D = 4$
space-time a chiral multiplet consists of a complex scalar field $z(x)$ and an 
anti-commuting chiral spinor $\psi_L(x)$. If  $z(x)$ takes values on the sphere
and carries a representation of spontaneously broken $SU(2)$, then these 
transformations commute with supersymmetry if we assign to the space-time
spinor the transformation rule of a vector over the target space (the sphere):
\be
\del_g \psi_L  = \der \xi_g(z)\, \psi_L = \lh i \vf + (\vthg - i \eta) z \rh \psi_L 
 =  2 f_g(z)  \psi_L.
\label{7.1}
\ee
Three remarks are in order: \\
- $\psi_L(x)$ is a fluctuating field over space-time, but a constant section of a
contravariant vector bundle over the sphere: it is independent of $z$, and therefore 
its derivatives w.r.t.\ $z$ vanish. \\
- Under supersymmetry the scalar field transforms into the chiral spinor $\psi_L$;
the transformation (\ref{7.1}) is precisely such, that supersymmetry and $SU(2)$ commute:
\be 
\del_Q z = \bar{\eps}_R \psi_L \hs{1} \Rightarrow \hs{1}
\left[ \del_g, \del_Q \right] z = 0. 
\label{7.2}
\ee 
- For the special case of the sphere it happens, that a vector transforms as a
line bundle of weight $2$; this is not true for general K\"{a}hler manifolds. 

The whole construction is equivalent to starting with a chiral superfield
\be
\Fg(x,\thg) = z(x) + \bar{\thg}_R \psi_L(x) + \bar{\thg}_R \thg_L H(x) +  ...,
\label{7.3}
\ee
where $\thg_L$ is an anticommuting spinor co-ordinate, and $H$ is a complex auxiliary
scalar field. Under $SU(2)$ this superfield transforms in the non-linear representation (\ref{3.1}):
\be
\del \Fg = \xi_g[\Fg] = i \vf\, \Fg + \frac{\vthg}{2} \lh 1 + \Fg^2 \rh + \frac{i\eta}{2}
 \lh 1 - \Fg^2 \rh.
\label{7.4}
\ee
The action of $SU(2)$ on the component fields can then be read off from a comparison 
of terms with the same $\thg
$ dependence on both sides of the equation. 
The construction of an invariant action for the chiral multiplet is
straightforward: first construct the real superfield-valued K\"{a}hler potential 
\be
K(\bFg, \Fg) = \ln \lh 1 + \bFg \Fg \rh;
\label{7.5}
\ee
then take the superspace integral ($D$-term) of this real superfield:
\be
S = \int d^4x \int d^2 \thg_L \int d^2 \thg_R\, K(\bFg, \Fg).
\label{7.6}
\ee
Written out in space-time components and eliminating the auxiliary fields
one obtains
\be 
S = - \int d^4x\, \lh \frac{\der \bz \cdot \der z}{(1 + \bz z)^2}
 + \frac{\bps_L \stackrel{\leftrightarrow}{\sDer} \ps_L}{(1 + \bz z)^2} 
 - \frac{(\bps_L \gam_{\mu} \ps_L)^2}{(1 + \bz z)^4} \rh.
\label{7.7}
\ee
Here the covariant derivative of the spinor field is defined as the pull-back of the
K\"{a}hler connection on the sphere:
\be 
\sDer \ps_L = \lh \sder  + \sder z\, \Gam_{zz}^{\;\;\;z} \rh \ps_L
 = \lh \sder - \frac{2\bz \sder z}{(1 + \bz z)^2} \rh \ps_L.
\label{7.8}
\ee
The first purely bosonic term of this action is precisely that of the non-linear $\sg$-model
on the sphere, cf.\ eq.(\ref{5.1}). The fermionic terms are dictated completely by the 
invariance of the action (modulo total divergences) under supersymmetry transformations
of the type (\ref{7.2}). 

\section{Anomalies} 

The action (\ref{7.7}) is invariant under both supersymmetry and $SU(2)$; now 
$SU(2)$, in particular its linear subgroup $U(1)$, acts non-trivially on the chiral
spinor $\psi_L$, cf.\ eq.(\ref{7.1}).  It is well-known that such a group action on chiral 
fermions is anomalous in the quantum theory: although the action is invariant, for a
single chiral fermion there is no invariant path-integral measure. The discussion can 
be simplified by redefining the fermion field:
\be
\chi_L = e^{-K(\bz, z)}\, \psi_L = \frac{\psi_L}{1 + \bz z}.
\label{8.1}
\ee
In terms of this field the action (\ref{7.7}) takes the form  
\be
S = - \int d^4x \lh \frac{\der \bz \cdot \der z}{(1 + \bz z)^2} + 
 \bar{\chi}_L \stackrel{\leftrightarrow}{\sDer} \chi_L - \lh \bar{\chi}_L \gam_{\mu} \chi_L \rh^2 \rh,
\label{8.2}
\ee
with the covariant derivative
\be
\sDer \chi_L = \lh \sder - i V\Slashed \rh \chi_L, \hs{2} 
 V_{\mu} = - i\, \frac{\bz  \stackrel{\leftrightarrow}{\der}_{\mu} z}{1 + \bz z}.
\label{8.3}
\ee
By construction this connection renders the derivative covariant under field-dependent 
$U(1)$ transformations
\be 
(\chi_L)_g = e^{f_g(z) - \bar{f}_g(\bz)}\, \chi_L \approx (1 + f_g(z) - \bar{f}_g(\bz)) \chi_L,
\label{8.4}
\ee
which is the equivalent of the field-dependent $SU(2)$ transformation of $\psi_L$, eq.(\ref{7.1}).

As the kinetic fermion term in (\ref{8.2}) is of the standard form, the usual triangle
anomaly calculation applies, and the $U(1)$ current, as well as its parent $SU(2)$ currents,
are not conserved. The quantum theory then is inconsistent \ct{moor-nel}-\ct{buch-lerch}. 
One way to repair this situation is to introduce another chiral multiplet, which we refer to
as `matter multiplet', 
\be
A(x, \thg) = a(x) + \bar{\thg}_R \fg_L(x) + \bar{\thg}_R \thg_L N(x) + ...
\label{8.5}
\ee
transforming as a holomorphic line bundle of weight $-2$:
\be
A_g = e^{-2f_g(\Fg)}\, A.
\label{8.6}
\ee
Note, that the holomorphicity is important to guarantee the chiral superfield nature of the 
transformed superfield.

For the chiral superfield $A$ one can construct an invariant action from the superspace
expression 
\be
\cL_{matter} = e^{2K(\bFg, \Fg)}\, \bar{A} A,
\label{8.7}
\ee
integrated over all of superspace \ct{sgn-nyaw-ricc-vh}. If one redefines the chiral
fermion $\fg_L$ by  
\be
\eta_L = e^{K(\bz,z)}  \fg_L = (1 + \bz z) \fg_L,
\label{8.8}
\ee
it has the opposite $U(1)$ transformation property compared to the quasi-Goldstone fermion
$\chi_L$:
\be
(\eta_L)_g = (1 - f_g(z) + \bar{f}_g(\bz)+\, {\cal O}(a))\, \eta_L. 
\label{8.8.1}
\ee 
In the  $U(1)$-invariant vacuum (i.e., $\langle a \rangle = 0$),  the kinetic terms of this fermion in the
action (\ref{8.7}) then become 
\be
\Del \cL_{matter} = \bar{\eta}_L \stackrel{\leftrightarrow}{\sDer} \eta_L +..., \hs{2}
\sDer \eta_L = \lh \sder + i V\Slashed \rh \eta_L.
\label{8.9}
\ee
As the effective $U(1)$ charge of the matter fermion $\eta_L$ is opposite to that of the 
quasi-Goldstone fermion $\chi_L$, their triangle anomalies cancel. The mechanism
explained here to eliminate anomalies in supersymmetric theories on K\"{a}hler manifolds
using holomorphic line-bundle representations of symmetry groups works quite generally
\ct{sgn-vh1}. 

\section{Gauging of internal symmetries}

Let me summarize the results of the previous sections. The supersymmetric model on
the coset $SU(2)/U(1)$, defined by the two chiral superfields $(\Fg, A)$ and the K\"{a}hler
potential 
\be
K[\bFg, \Fg; \bA, A] = \ln (1 + \bFg \Fg) + (1 + \bFg \Fg)^2 \bA A,
\label{9.1}
\ee
is invariant under the non-linear $SU(2)$ transformations
\be
\ba{l}
\dsp{ \del \Fg = \eps + i \vf\, \Fg + \beps\, \Fg^2, \hs{2} 
\del A = - i \vf A - 2 \beps\, \Fg A, }\\
 \\
\dsp{ \eps = \frac{1}{2} \lh \vthg + i \eta \rh, \hs{2} \beps = \frac{1}{2} \lh \vthg - i\eta \rh. }
\ea 
\label{9.2}
\ee
As the two chiral fermions in the model carry opposite charges, the internal symmetry is
free of anomalies. This allows us to promote the model to a consistent quantum field theory 
e.g.\ using path-integral quantization. Of course, the model is not renormalizable, hence it
must be regarded as an effective quantum field theory, decribing the physical degrees 
of freedom in a limited range of energies/distance scales.

A second important benefit from the absence of anomalies and the resulting current
conservation is, that the internal symmetry can be gauged consistently. Because of the 
interplay with supersymmetry, gauging the $SU(2)$ symmetry involves several steps.
The first step is to extend all derivatives to covariant derivatives, introducing gauge fields
$(W^+_{\mu}, A_{\mu}, W^-_{\mu})$ for the transformations parametrized by $(\eps, \vf, \beps)$,
respectively. For example, the covariant derivatives on the scalars act as
\be
D_{\mu} z = \der_{\mu} z - g (W^+_{\mu} + i A_{\mu}z + W^-_{\mu} z^2), \hs{1}
D_{\mu} a = \der_{\mu} a + g(i A_{\mu} + 2 W^-_{\mu} z)\, a.
\label{9.3}
\ee
In the second step, one adds associated Yukawa couplings of the complex scalars $(z, a)$ 
to the fermions $(\psi_L, \fg_L)$ and the gauginos $(\lb^+, \lb, \lb^-)$, the superpartners of the
gauge fields. And finally one has to add a potential which results from eliminating the auxiliary 
fields $(D^+, D, D^-)$ associated with the gauge fields by supersymmetry. We will not present
the full action here; it can been found in ref.\ct{sgn-nyaw-ricc-vh}. However, the $D$-term 
potential is of special interest, and we discuss it in some detail.

I first return to eq.(\ref{3.8}) defining the holomorphic Killing vectors $\xi^i(z)$. This equation
implies, that at least locally for any Killing vector there exists a real function $M(\bz,z)$ such 
that
\be 
\xi_{\ui} = g_{i\ui}\, \xi^i = -i \dd{M}{\bz^{\ui}}, \hs{2}
\bxi_{i} = g_{i\ui}\, \bxi^{\ui} = i \dd{M}{z^i}. 
\label{9.4}
\ee
Actually an explicit construction of these Killing potentials exists. Consider the Killing
vector $\xi_g(z)$ associated with the element $g$ of the group of invariances of the
line element. We know that under this transformation the K\"{a}hler potential behaves as
\be 
\del_g K = \xi_g^i\, \dd{K}{z^i} + \bxi_g^{\ui}\, \dd{K}{\bz^{\ui}} = f_g(z) + \bar{f}_g(\bz)
\label{9.5}
\ee
Now define $M_g(\bz, z)$ by  
\be
- iM_g = \xi_g^i\, \dd{K}{z^i} - f_g(z) = - \bxi_g^{\ui}\, \dd{K}{\bz^{\ui}} + \bar{f}_g(\bz). 
\label{9.6}
\ee
As is obvious from the expressions on the r.h.s., the quantity $M_g$ is real. It is
straightforward to show, that $M_g$ is the Killing potential for $\xi_g^i(z)$; indeed, 
as both $\xi^i_g(z)$ and $f_g(z)$ are holomorphic,
\be 
- i \dd{M_g}{\bz^{\ui}} = \xi_g^i\, \dd{^2K}{\bz^{\ui} \der z^i} = g_{i\ui}\, \xi_g^i = \xi_{g\,\ui}.
\label{9.7}
\ee
Applying this construction to the coset model on $SU(2)/U(1)$ we find the explicit 
expressions
\be 
- i M_g(\bFg, \Fg; \bA, A) = \frac{ -i \vf\, (1  - \bFg \Fg) + 2(\eps \bFg - \beps \Fg)}{
 1 + \bFg \Fg}  \lh \frac{1}{2}\, + \lh 1 + \bFg \Fg \rh^2 \bA A \rh.
\label{9.8}
\ee
Note, that if one computes the gradient of these potentials: $ \der M_g/\der \bFg$,
$\der M_g/\der \bA$, one does not recover directly the Killing vectors represented
by eqs.(\ref{9.2}): there is still a metric factor
\be 
G_{I\uI} = \lh \ba{cc} \dd{^2K}{\bFg \der \Fg} & \dd{^2K}{\bFg \der A} \\
                                              \\
                                  \dd{^2K}{\bA \der \Fg} & \dd{^2K}{\bA \der A} \ea \rh,
\label{9.9}
\ee
to take into account. 

Returning to the subject of the $D$-term potential, it can be shown quite generally that 
elimination of the auxilary $D$-fields leads to a scalar potential
\be 
V_D(\bz, z; \bar{a}, a) = \frac{g^2}{2}\, D_a^2 = \frac{g^2}{2}\, M_a^2(\bz, z;\bar{a}, a),
\label{9.10}
\ee
where the sum is over all independent components of the Lie-algebra of isometries, 
labeled by $a$. For the case of the coset $SU(2)/U(1)$ this becomes explicitly
\be 
V_D = \frac{g^2}{2} \dd{M_g}{\eps} \dd{M_g}{\beps} +  \frac{g^2}{2} \lh \dd{M_g}{\vf} \rh^2
 = \frac{g^2}{2} \lh \frac{1}{2}\, + (1 + \bz z)^2 \bar{a} a \rh^2. 
\label{9.11}
\ee
From the explicit form of the potential we learn two important physical facts: \\
- The potential is positive definite, with a minimum at $V_D = g^2/8$; this implies
that supersymmetry is spontaneously broken. \\
- The minimum is reached for $z = a  =0$; hence the linear $U(1)$ symmetry 
is not spontaneously broken, and the $U(1)$ gauge field $A_{\mu}$ remains
massless. \\
We observe at the same time, that the charged vector bosons $W^{\pm}_{\mu}$
become massive. This is most easily seen by going to the unitary gauge, in which
$\bz =z = 0$; then the covariant derivative of $z$, eq.(\ref{9.3}), reduces to  $-g W_{\mu}^+$,
and the former kinetic terms for the Goldstone bosons become
\be
\cL_{kin}(\bz, z) = - g^2 W^+ \cdot W^-.
\label{9.12}
\ee
I refer to \ct{sgn-nyaw-ricc-vh} for details.

\section{Extensions to larger coset manifolds}

The action of a supersymmetric field theory in 4-$D$ space-time constructed
from chiral supermultiplets $(z^i, \psi_L^i)$, $i = 1, ..., r$,  is defined by two functions
of the superfields: the real K\"{a}hler potential $K(\bFg, \Fg)$, and the holomorphic
superpotential $W(\Fg)$. In components the action reads
\be
\ba{lll} 
\cL & = & \dsp{ - g_{i\ui} \lh \der \bz^{\,\ui} \cdot \der z^i + \frac{1}{2}\, \bar{\ps}_L^{\ui}
 \stackrel{\leftrightarrow}{\sDer} \ps_L^i \rh - g^{\ui i} \bar{W}_{;\ui} W_{;i} }\\
 & & \\
 & & \dsp{ +\, W_{;ij}\, \bar{\ps}^{i}_R \ps_L^{j} + \bar{W}_{;\ui \uj}\, \bar{\ps}_L^{\ui} \ps_R^{\uj} 
 + \frac{1}{8}\, R_{i\ui j\uj}\, \bar{\ps}_L^{\ui} \gam^{\mu} \ps_L^i\,
 \bar{\ps}_R^{\uj} \gam_{\mu} \ps_R^{\ui}.  }
\ea
\label{e.1}
\ee
Here $g_{i\ui} = K_{,i\ui}$ is the K\"{a}hler metric, $\sDer$ the covariant derivative 
on the K\"{a}hler manifold and $R_{i\ui j\uj}$ the corresponding curvature tensor.
There are many cosets of K\"{a}hler type of interest for particle physics
phenomenology; these include the Grasmannian models on $SU(N+M)/ S[U(N) \times
U(M)]$,  such as the GUT-like model $SU(5)/SU(3) \times SU(2) \times U(1)$. 

To reproduce the particle content of the standard model, as summarized in table 1,  the 
internal symmetries of such a model must be promoted to local symmetries, by coupling
to appropriate non-abelian vector multiplets; this is possible only if the symmetries are 
non-anomalous. Like in the model on $SU(2)/U(1)$, this is generically not the case. The
model on $SU(5)/SU(3) \times SU(2) \times U(1)$ provides a typical example. The 
Goldstone superfields in this model are $\Fg^{\ag r}$, where $\ag = (1,2)$ is an $SU(2)$
index and $r = (1,2,3)$ an $SU(3)$ index; thus they transform as a doublet-triplet of $SU(2) 
\times SU(3)$.  In particular, the fermion components carry the quantum numbers of
left-handed quark doublets. By itself this set of chiral fermions forms an anomalous 
representation of $SU(3) \times SU(2) \times U(1)$.

As in the $SU(2)/U(1)$ model,  the anomalies can be canceled by incorporating additional
supermultiplets. Actually the standard model suggests how to do this: introduce 
chiral superfields with the $SU(3) \times SU(2)$ quantum numbers of lepton doublets, plus
superfields containing anti-quark and anti-lepton $SU(2)$ singlets. Indeed, this suffices to 
guarantee the absence of anomalies of the full parent $SU(5)$ symmetry, provided the chiral
fermions have the correct $U(1)$ hypercharges. Observe, that to obtain full agreement with 
the spontaneously broken standard model including at least one family of quarks and leptons,
one needs to introduce a Higgs and an anti-Higgs doublet with opposite hypercharges as well; 
with this hypercharge assignment the higgsinos do not cause new anomalies. Details of this
model can be found in \ct{sgn-vh2}.

To realize the correct hypercharge assignments of the additional superfields is not trivial. 
For example,  the left-handed anti-electron is an $SU(2)$ and $SU(3)$ singlet, suggesting
that it should be part of a superfield $\Fg$ which is a singlet under $SU(3) \times SU(2) 
\times U(1)$; however, its true $U(1)$ hypercharge is $+1$. Such a hypercharge assignment
can be realized by promoting $\Fg$ to a line-bundle representation of $SU(5)$, generalizing 
the construction I presented for $SU(2)/U(1)$. The required line-bundle representation
actually exists, not only for the left-handed anti-electron, but for all chiral superfields containing 
standard-model fermions and Higgs doublets \ct{sgn-vh2}. Thus line bundle representations
play a crucial role in eliminating anomalies, not only in the simple $SU(2)/U(1)$ model, but 
also in the larger models which are more interesting from the phenomenological point of view.

Once the anomalies have been canceled, the $SU(5)$ symmetry can be gauged in a way
that respects the supersymmetry of the action. In broad outline the following happens: \\
- First, a set of 24 vector multiplets $(A_{\mu}^a, \lb^a, D^a)$ is introduced
as required by local $SU(5)$ symmetry. \\
- Gauging the action whilst preserving supersymmetry gives rise to covariant derivatives,
accompanied  by Yukawa couplings and a $D$-term potential,  defined in terms of the
Killing potentials for the $SU(5)$ isometries. \\
- The vector bosons corresponding to broken directions in $SU(5)$ become massive, but
the vector bosons corresponding to the linear $SU(3) \times SU(2) \times U(1)$
subgroup remain massless, at least as long as there are no vacuum expectation
values for the Higgs doublets. \\
- The Goldstone bosons $z^i$ disappear from the spectrum of physical states;
this is manifest in the unitary gauge $z^i = 0$, which can always be realized if the
parameters of the non-linear $SU(5)$ transformations become space-time dependent. 
As a result, in these supersymmetric models only the quarks and leptons which are
part of the additional matter superfields have scalar partners; the quark doublet-triplet
has no physical superpartners left. This shows, that the model exhibits strong spontaneous
supersymmetry breaking. \\
- Like the vector bosons, the gauginos split into linear representations of the stability
group $SU(3) \times SU(2) \times U(1)$:
\be
{\bf 24} \hs{1} \rightarrow \hs{1} ({\bf 8}, {\bf 1}) + ( {\bf 1}, {\bf 3}) + ({\bf 1}, {\bf 1})
 + ({\bf 3},{\bf 2})_R + ({\bf 3}, {\bf 2})_L.
\label{9.13}
\ee
The fate of these fermions is quite interesting; first, there are three sets of Majorana gauginos 
with the same quantum numbers as the massless vector bosons, as expected. Then
there is a righthanded doublet-triplet which (in the unitary gauge) combines with the
quasi-Goldstone fermions $\psi_L$ to form a doublet of massive quarks, with masses
of the order of the scale at which $SU(5)$ is broken. And finally, there remains a 
massless doublet-triplet representing the physical quark doublets in the gauged
$\sg$-model, without physical scalar superpartners. \\
- The $D$-term potential again takes the form
\be
V_D = \frac{g^2}{2}\, M_a^2,
\label{9.14}
\ee
where $g$ is the $SU(5)$ coupling constant, and the $M_a$ are the Killing potentials of the
$SU(5)$ transformations. In this expression one may take $\bz^{\ui} = z^i = 0$ (unitary gauge),
i.e.\ the potential depends only on the physical scalars. It is positive definite, confirming that
supersymmetry is broken spontaneously.

The scenario sketched here for the particular model based on the grassmannian coset
 $SU(5)/SU(3) \times SU(2) \times U(1)$ is representative for a class of models incorporating
standard-model like low-energy sectors, e.g.\  the models on $SO(10)/U(5)$ and
$E_6/SO(10) \times U(1)$ \ct{sgn-nyaw-vh}; of course the details differ. The coupling of
these models to supergravity has been discussed in \ct{sgn-vh2}.

\section{Conclusions}

In these notes I have explained the basic elements of K\"{a}hler geometry
and its application to the construction of 4-D supersymmetric field theories.
In particular I have shown that there exist consistent supersymmetric models with
the fermion content and  gauge interactions of the standard model, but different
from the minimal supersymmetric standard model (MSSM).

In these supersymmetric coset models all or some of the quarks and leptons are
quasi-Goldstone fermions, or an equivalent set of unpaired chiral gauginos, lacking
physical scalar partners. This strong form of spontaneous supersymmetry breaking
clearly distinguishes the physical content of these models from the MSSM or standard
supersymmetric GUTs. Of course, this difference affects scenarios of gauge unification
and the role of supersymmetry in the solution of the hierarchy problem, although to
what extent is presently not clear.

In addition, questions arise as to the possibility of breaking the stability group $H$,
which is to be chosen as the standard-model group or a GUT group. In some models
we have found that the scalar potential drives the scalar fields to a singular point
of the kinetic terms \ct{sgn-nyaw-vh, sgn-nyaw-ricc-vh}. This may point to a
strong coupling regime, possibly with symmetry restauration. Clearly, not all of the
physics described by these models is as yet understood.

With the principles of constructing a class of consistent models at hand, the next step
is to address physical applications. This part of the work is only about to begin.

\np

\end{document}